\documentclass[aps,prd,nofootinbib,amsmath,amssymb,showpacs,superscriptaddress,twocolumn]{revtex4}
\usepackage{txfonts}
\usepackage{graphicx}
\usepackage{dcolumn}
\usepackage{bm}
\usepackage{amssymb}
\usepackage{latexsym}
\usepackage[colorlinks, linkcolor=blue, citecolor=blue, urlcolor=blue]{hyperref}

\newcommand{\be}{\begin{equation}}
\newcommand{\ee}{\end{equation}}
\newcommand{\bq}{\begin{eqnarray}}
\newcommand{\eq}{\end{eqnarray}}

\begin{document}

\title{Probing the dynamics of dark energy with novel parametrizations}

\author{Jing-Zhe Ma}
\affiliation{Department of Physics, College of Sciences,
Northeastern University, Shenyang 110004, China}
\author{Xin Zhang}
\affiliation{Department of Physics, College of Sciences,
Northeastern University, Shenyang 110004, China} \affiliation{Center
for High Energy Physics, Peking University, Beijing 100080, China}

\begin{abstract}
We point out that the CPL parametrization has a problem that the
equation of state $w(z)$ diverges in the far future, so that this
model can only properly describe the past evolution but cannot
depict the future evolution. To overcome such a difficulty, in this
Letter we propose two novel parametrizations for dark energy, the
logarithm form $w(z)=w_0+w_1({\ln (2+z)\over 1+z}-\ln2)$ and the
oscillating form $w(z)=w_0+w_1({\sin(1+z)\over 1+z}-\sin(1))$,
successfully avoiding the future divergency problem in the CPL
parametrization, and use them to probe the dynamics of dark energy
in the whole evolutionary history. Our divergency-free
parametrizations are proven to be very successful in exploring the
dynamical evolution of dark energy and have powerful prediction
capability for the ultimate fate of the universe. Constraining the
CPL model and the new models with the current observational data, we
show that the new models are more favored. The features and the
predictions for the future evolution in the new models are discussed
in detail.
\end{abstract}

\pacs{95.36.+x, 98.80.Es, 98.80.-k}

\keywords{Dark energy; dynamical evolution; CPL parametrization;
novel parametrizations}

\maketitle

The properties of dark energy are mainly characterized by the
equation of state parameter (EOS), $w$. Extracting the information
of EOS of dark energy from observational data is very challenging
owing to the accuracy of current data and our ignorance of dark
energy. For probing the dynamical evolution of dark energy, under
such circumstance, one has to parameterize $w$ empirically, usually
using two or more free parameters. Among all the parametrization
forms of EOS, the Chevallier-Polarski-Linder (CPL) model~\cite{CPL}
is the most widely used one and has been explored extensively. The
form of the CPL parametrization is
\begin{equation}\label{eq4}
w(z)=w_0+w_1\frac{z}{1+z},
\end{equation}
where $z$ is the redshift, $w_0$ is the present-day value of the
EOS, and $w_1$ is the derivative of the EOS with respect to the
scale factor $a$. The direct motivation of proposing such a
parametrization form is to overcome the divergency feature of the
linear form $w(z)=w_0+w_1 z$ at high redshifts. Furthermore, as
Linder~\cite{CPL} suggested, the CPL parametrization has several
advantages, such as a manageable two-dimensional phase space, well
behaved and bounded behavior for high redshifts, high accuracy in
reconstructing many scalar field equations of state, simple physical
interpretation, etc.

However, we have to point out that there exists a problem in the CPL
model. The CPL model only explores the past expansion history
properly, but cannot describe the future evolution due to the fact
that $|w(z)|$ grows increasingly and finally encounters divergency
as $z$ approaches $-1$. Undoubtedly, this is a nonphysical feature.
Such a divergency problem prevents the CPL parametrization from
genuinely covering the scalar-field models as well as other
theoretical models.

To overcome the shortcoming of the CPL model, we are interested in
proposing novel parametrization forms of $w(z)$. The new
parametrizations will be contrived to exceed the CPL model entirely:
inheriting the advantages of the CPL model, avoiding the disastrous
divergency in the far future, and being more favored by the
observational data. In this Letter, we are devoted to exploring more
insightful parametrization forms for dark energy and probing the
dynamics of dark energy in light of the novel parametrizations.

The leading proposal we put forth for the EOS of dark energy is of
the form:
\begin{equation}\label{eq6}
w(z)=w_0+w_1\left(\frac{\ln(2+z)}{1+z}-\ln2\right),
\end{equation}
where $w_0$ also denotes the present-day value of $w(z)$, and $w_1$
is another parameter characterizing the evolution of $w(z)$. Note
that a minus $\ln2$ in the bracket is kept for maintaining $w_0$ to
be the current value of $w(z)$ and for easy comparison with the CPL
model. Obviously, this new parametrization has well behaved, bounded
behavior for both high redshifts and negative redshifts. Thanks to
the logarithm form in the parametrization, a finite value for $w(z)$
can be ensured, via the application of the L'Hospital's rule, in
both limiting cases, $z\rightarrow\infty$ and $z\rightarrow -1$.
This is the reason why we introduce a logarithm form in the new
parametrization. For clearness, we list the values of $w(z)$ in the
limiting cases:
\begin{eqnarray}\label{eq7}
{w(z)} = \left\{\begin{array}{ll} w_0,\ \ \ \ \ \ \ \ \ \ &$for$\
z=0,\\\\
w_0-w_1\ln2,&$for$\ z\rightarrow+\infty,\\\\
w_0+w_1(1-\ln2),&$for$\ z\rightarrow-1.
\end{array}\right.
\end{eqnarray}
At low redshifts, the new form reduces to the linear one,
$w(z)\approx w_0+\tilde{w}_1z$, where $\tilde{w}_1=-(\ln2)w_1$. Of
course, one can also recast it at low redshifts as the CPL form,
$w(z)\approx w_0+\tilde{w}_1z/(1+z)$, where
$\tilde{w}_1=(1/2-\ln2)w_1$. Therefore, it is clear to see that the
new parametrization exhibits well-behaved feature for the dynamical
evolution of dark energy. Without doubt, such a two-parameter form
of EOS can genuinely cover many scalar-field models (including
quintom models with two scalar fields and/or with one field with
high derivatives) as well as other theoretical scenarios.

It is well-known that dark energy drives the cosmic acceleration
only at the late times ($z\sim 0.5$), whereas at the early times
dark energy can be totally neglected due to the extremely low
density compared to the matter or radiation component. Thus, one can
only justify that the EOS of dark energy is around $-1$ in the
recent epoch, but can tolerate more possibilities for the early-time
EOS of dark energy. For example, the EOS of dark energy might
exhibit oscillating feature during the evolution~\cite{oscide}. This
is a fascinating possibility, deserving a detailed investigation.
Based on this consideration, we further extend the above new
parametrization (\ref{eq6}) to an oscillating one, by replacing the
logarithm function with a sine function:
\begin{equation}\label{eq9}
w(z)=w_0+w_1\left(\frac{\sin(1+z)}{1+z}-\sin(1)\right).
\end{equation}
Such a replacement is rather reasonable, lying in the fact that the
two parametrizations roughly coincide in the recent epoch (low
redshifts), since $\sin(1)\approx \ln2$ and $\cos(1)\approx 1/2$
[note that $\sin(1)\approx0.841$, $\ln2\approx 0.693$, and
$\cos(1)\approx0.540$]. Hence, the parametrization (\ref{eq9})
describes the same behavior as the logarithm form (\ref{eq6}) at low
redshifts, but exhibits oscillating feature from a long term point
of view. Also, we list the values of $w(z)$ in the following
limiting cases:
\begin{eqnarray}\label{eq10}
{w(z)} = \left\{\begin{array}{ll} w_0,\ \ \ \ \ \ \ \ \ \ &$for$\
z=0,\\\\
w_0-w_1\sin(1),&$for$\ z\rightarrow+\infty,\\\\
w_0+w_1(1-\sin(1)),&$for$\ z\rightarrow-1.
\end{array}\right.
\end{eqnarray}
We find that the two parametrizations, (\ref{eq6}) and (\ref{eq9}),
also roughly coincide in the limiting cases, $z\rightarrow\infty$
and $z\rightarrow -1$.

In what follows, we shall explore the dynamical evolution of dark
energy via the CPL parametrization and the new parametrizations. For
convenience, we call the new parametrizations the logarithm form and
oscillating form, respectively, hereafter. Since our aim is to probe
the dynamics of dark energy, we should try to avoid other indirect
factors weakening the observational limits on the EOS; thus we
assume a flat universe, $\Omega_k=0$, consistent with the
inflationary cosmology. From the Friedmann equation, the Hubble
expansion rate can be written as
\begin{equation}\label{eq1}
H(z)=H_0\left[\Omega_{m}(1+z)^3+\Omega_{r}(1+z)^4+(1-\Omega_m-\Omega_r)f(z)\right]^{1/2},
\end{equation}
where $\Omega_r=\Omega_\gamma(1 + 0.2271N_{\rm eff})$, with
$\Omega_\gamma=2.469\times 10^{-5}h^{-2}$ for $T_{\rm cmb}=2.725$ K,
$N_{\rm eff}$ the effective number of neutrino species (in this
Letter we take its standard value, 3.04 \cite{WMAP7}), and
$f(z)=\exp[3\int_0^z dz'(1+w(z'))/(1+z')]$.

For constraining $w(z)$, we use the current observational data from
the type Ia supernovae (SN), the baryon acoustic oscillations (BAO),
and the cosmic microwave background (CMB). Such a combination of
data sets is the most widely used one, sufficiently satisfying our
aim of testing the new parametrizations and making a comparison. Of
course, one can also add other data sets such as gamma-ray bursts,
$H(z)$, and so on, but we feel that this is not necessary for our
present aim and leave a more sophisticated analysis to a future work
with different goal.

For the SN data, we use the $557$ Union2 data compiled in
Ref.~\cite{Amanullah:2010vv}. The theoretical distance modulus is
defined as
 \be\label{eq12}
   \mu_{th}(z_i)\equiv5\log_{10} D_L(z_i)+\mu_0,
 \ee
where $\mu_0\equiv42.38-5\log_{10} h$ with $h$ the Hubble constant
$H_0$ in units of $100$ km/s/Mpc, and the Hubble-free luminosity
distance
 \be\label{eq13}
   D_L(z)=(1+z)\int_0^z \frac{dz'}{E(z';{\bm \theta})},
 \ee
where $E\equiv H/H_0$, and ${\bm\theta}$ denotes the model
parameters. Correspondingly, the $\chi^2$ function for the $557$
Union2 SN data is given by
 \be\label{eq14}
   \chi^2_{SN}({\bm\theta})=\sum\limits_{i=1}^{557}\frac{\left[\mu_{obs}(z_i)-\mu_{th}(z_i)\right]^2}{\sigma^2(z_i)},
 \ee
where $\sigma$ is the corresponding $1\sigma$ error of distance
modulus for each supernova. The parameter $\mu_0$ is a nuisance
parameter but it is independent of the data points. Following
Ref.~\cite{Nesseris:2005ur}, the minimization with respect to
$\mu_0$ can be made trivially by expanding $\chi^2$ of
Eq.~(\ref{eq12}) with respect to $\mu_0$ as
 \be\label{eq13}
   \chi^2_{SN}({\bm\theta})=A-2\mu_0 B+\mu_0^2 C,
 \ee
where
 $$A({\bm\theta})=\sum\limits_{i=1}^{557}\frac{\left[\mu_{obs}(z_i)-\mu_{th}(z_i;\mu_0=0,{\bm\theta})\right]^2}{\sigma_{\mu_{obs}}^2(z_i)},$$
 $$B({\bm\theta})=\sum\limits_{i=1}^{557}\frac{\mu_{obs}(z_i)-\mu_{th}(z_i;\mu_0=0,{\bm\theta})}{\sigma_{\mu_{obs}}^2(z_i)},$$
 $$C=\sum\limits_{i=1}^{557}\frac{1}{\sigma_{\mu_{obs}}^2(z_i)}.$$
Evidently, Eq.~(\ref{eq13}) has a minimum for $\mu_0=B/C$ at
 \be\label{eq14}
   \tilde{\chi}^2_{SN}({\bm\theta})=A({\bm\theta})-\frac{B({\bm\theta})^2}{C}.
 \ee
Since $\chi^2_{SN,\,min}=\tilde{\chi}^2_{SN,\,min}$, instead
minimizing $\chi_{SN}^2$ we will minimize $\tilde{\chi}^2_{SN}$
which is independent of the nuisance parameter $\mu_0$.

For the BAO measurement, we use the data from SDSS DR7
\cite{Percival:2009xn}. The distance ratio ($d_z$) at $z=0.2$ and
$z=0.35$ are
\begin{equation}
d_{0.2}=\frac{r_{s}(z_{d})}{D_{V}(0.2)},~~
d_{0.35}=\frac{r_{s}(z_{d})}{D_{V}(0.35)},
\end{equation}
where $r_{s}(z_{d})$ is the comoving sound horizon at the baryon
drag epoch \cite{Eisenstein:1997ik}, and
\begin{equation}
D_{V}(z)=\left[\left(\int_{0}^{z}\frac{dz'}{H(z')}\right)^{2}\frac{z}{H(z)}\right]^{1/3}
\end{equation}
encodes the visual distortion of a spherical object due to the non
Euclidianity of a FRW spacetime. The inverse covariance matrix of
BAO is
\begin{eqnarray} (C^{-1}_{BAO}) & = & \left(\begin{array}{ccc}
30124 & -17227 \\
-17227 & 86977\end{array}\right).\end{eqnarray} The $\chi^2$
function of the BAO data is constructed as: \be\label{eq19}
\chi_{BAO}^2=(d_i^{th}-d_i^{obs})(C_{BAO}^{-1})_{ij}(d_j^{th}-d_j^{obs}),
\ee where $d_i=(d_{0.2}, d_{0.35})$ is a vector, and the BAO data we
use are $d_{0.2}=0.1905$ and $d_{0.35}=0.1097$.

The CMB is sensitive to the distance to the decoupling epoch via the
locations of peaks and troughs of the acoustic oscillations. In this
Letter, we employ the ``WMAP distance priors'' given by the
seven-year WMAP observations \cite{WMAP7}. This includes the
``acoustic scale'' $l_A$, the ``shift parameter'' $R$, and the
redshift of the decoupling epoch of photons $z_*$. The acoustic
scale $l_A$ describes the distance ratio $D_A(z_*)/r_s(z_*)$,
defined as
\begin{equation}
\label{ladefeq} l_A\equiv (1+z_*){\pi D_A(z_*)\over r_s(z_*)},
\end{equation}
where a factor of $(1+z_*)$ arises because $D_A(z_*)$ is the proper
angular diameter distance, whereas $r_s(z_*)$ is the comoving sound
horizon at $z_*$. The fitting formula of $r_s(z)$ is given by
\begin{equation}
r_s(z)=\frac{1} {\sqrt{3}}  \int_0^{1/(1+z)}  \frac{ da } { a^2H(a)
\sqrt{1+(3\Omega_{b}/4\Omega_{\gamma})a} },
\end{equation}
where $\Omega_{b}$ and $\Omega_{\gamma}$ are the present-day baryon
and photon density parameters, respectively. In this Letter, we fix
$\Omega_{\gamma}=2.469\times10^{-5}h^{-2}$ and $\Omega_{b}=0.02246
h^{-2}$ given by the seven-year WMAP observations \cite{WMAP7}. We
use the fitting function of $z_*$ proposed by Hu and Sugiyama
\cite{Hu:1995en}:
\begin{equation}
\label{zstareq} z_*=1048[1+0.00124(\Omega_b
h^2)^{-0.738}][1+g_1(\Omega_m h^2)^{g_2}],
\end{equation}
where
\begin{equation}
g_1=\frac{0.0783(\Omega_b h^2)^{-0.238}}{1+39.5(\Omega_b
h^2)^{0.763}},\quad g_2=\frac{0.560}{1+21.1(\Omega_b h^2)^{1.81}}.
\end{equation}
The shift parameter $R$ is responsible for the distance ratio
$D_A(z_*)/H^{-1}(z_*)$, given by \cite{Bond97}
\begin{equation}
\label{shift} R(z_*)\equiv \sqrt{\Omega_m H_0^2}(1+z_*)D_A(z_*).
\end{equation}
Following Ref.~\cite{WMAP7}, we use the prescription for using the
WMAP distance priors. Thus, the $\chi^2$ function for the CMB data
is
\begin{equation}
\chi_{CMB}^2=(x^{th}_i-x^{obs}_i)(C_{CMB}^{-1})_{ij}(x^{th}_j-x^{obs}_j),\label{chicmb}
\end{equation}
where $x_i=(l_A, R, z_*)$ is a vector, and $(C_{CMB}^{-1})_{ij}$ is
the inverse covariance matrix. The seven-year WMAP observations
\cite{WMAP7} give the maximum likelihood values: $l_A(z_*)=302.09$,
$R(z_*)=1.725$, and $z_*=1091.3$. The inverse covariance matrix is
also given in Ref.~\cite{WMAP7}:
\begin{equation}
(C_{CMB}^{-1})=\left(
  \begin{array}{ccc}
    2.305 & 29.698 & -1.333 \\
    29.698& 6825.27 & -113.180 \\
    -1.333& -113.180 &  3.414 \\
  \end{array}
\right).
\end{equation}

Since the SN, BAO and CMB are effectively independent measurements,
we can combine the data sets by simply adding together the $\chi^2$
functions. Thus, we have
\begin{equation}
\chi^2=\tilde{\chi}^2_{SN}+\chi^2_{BAO}+\chi^2_{CMB}.
\end{equation}
Note that $\tilde{\chi}^2_{SN}$ is free of $h$, while $\chi^2_{BAO}$
and $\chi^2_{CMB}$ are still relevant to $h$.

The three models have the same free model parameters, namely,
${\bm\theta}=\{\Omega_{m},~w_0,~w_1,~h\}$. According to the joint
data analysis, we obtain the best-fit parameters and the
corresponding $\chi^2_{min}$. The best-fit, $1\sigma$ and $2\sigma$
values of the parameters with $\chi^2_{min}$ of the three models are
all presented in Table~\ref{Table1}.

\begin{table*} \caption{The fitting values for the CPL, logarithm (Log) and oscillating (Sin) models. }
\begin{center}
\label{Table1}
\begin{tabular}{ccccccc}
  \hline\hline
  Model   &                         $\Omega_{m}$                               &                                                   $w_0$                                    &                                               $w_1$                                           &                                              $h$                                    &                 $\chi^2_{min}$        \\
  \hline
  CPL                ~&~$0.279^{+0.032}_{-0.028}\left(1\sigma\right)^{+0.047}_{-0.038}\left(2\sigma\right)$ ~~&~~$-1.066^{+0.267}_{-0.232}\left(1\sigma\right)^{+0.410}_{-0.332}\left(2\sigma\right)$~~&~~$0.261^{+0.904}_{-1.585}\left(1\sigma\right)^{+1.127}_{-2.608}\left(2\sigma\right)$ ~~&~~$0.699^{+0.029}_{-0.034}\left(1\sigma\right)^{+0.041}_{-0.047}\left(2\sigma\right)$ ~&~ 544.186    ~~~  \\
  \hline
  Log                ~&~$0.280^{+0.032}_{-0.028}\left(1\sigma\right)^{+0.047}_{-0.040}\left(2\sigma\right)$ ~~&~~$-1.067^{+0.234}_{-0.155}\left(1\sigma\right)^{+0.343}_{-0.210}\left(2\sigma\right)$~~&~~$-1.049^{+5.706}_{-0.896}\left(1\sigma\right)^{+8.708}_{-1.018}\left(2\sigma\right)$~~&~~$0.697^{+0.031}_{-0.026}\left(1\sigma\right)^{+0.046}_{-0.036}\left(2\sigma\right)$ ~&~ 544.081    ~~~  \\
  \hline
  Sin                ~&~$0.280^{+0.033}_{-0.029}\left(1\sigma\right)^{+0.050}_{-0.040}\left(2\sigma\right)$ ~~&~~$-1.061^{+0.191}_{-0.172}\left(1\sigma\right)^{+0.291}_{-0.249}\left(2\sigma\right)$~~&~~$-0.041^{+2.194}_{-0.745}\left(1\sigma\right)^{+3.678}_{-0.935}\left(2\sigma\right)$~~&~~$0.695^{+0.034}_{-0.038}\left(1\sigma\right)^{+0.046}_{-0.054}\left(2\sigma\right)$ ~&~ 543.986    ~~~  \\
  \hline\hline
\end{tabular}
\end{center}
\end{table*}

\begin{figure*}[htbp]
\centering
\begin{center}
$\begin{array}{c@{\hspace{0.2in}}c} \multicolumn{1}{l}{\mbox{}} &
\multicolumn{1}{l}{\mbox{}} \\
\includegraphics[scale=0.68]{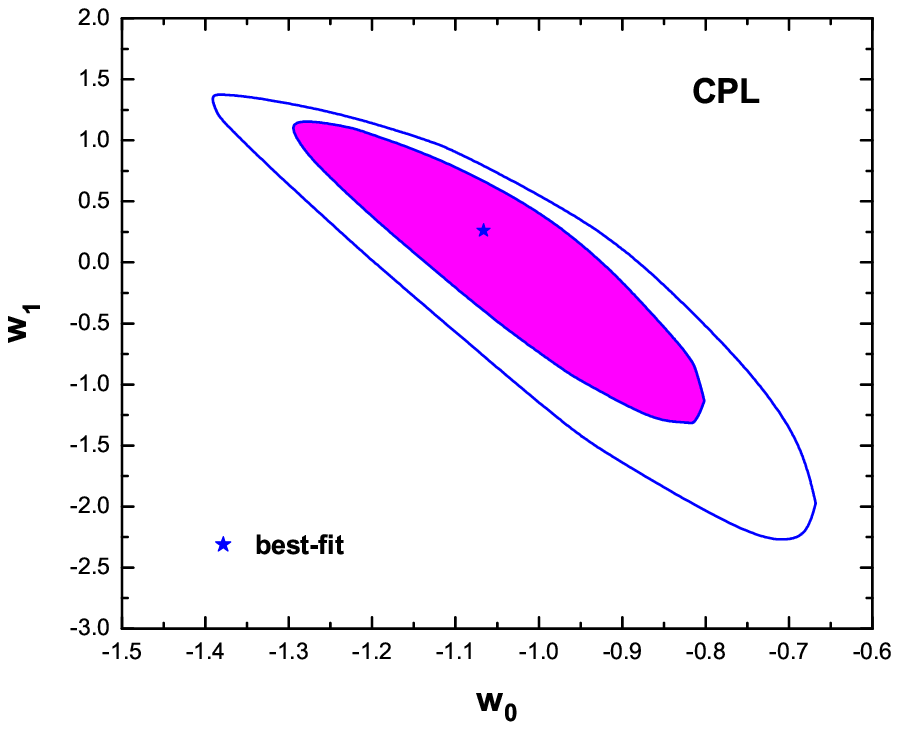} &\includegraphics[scale=0.68]{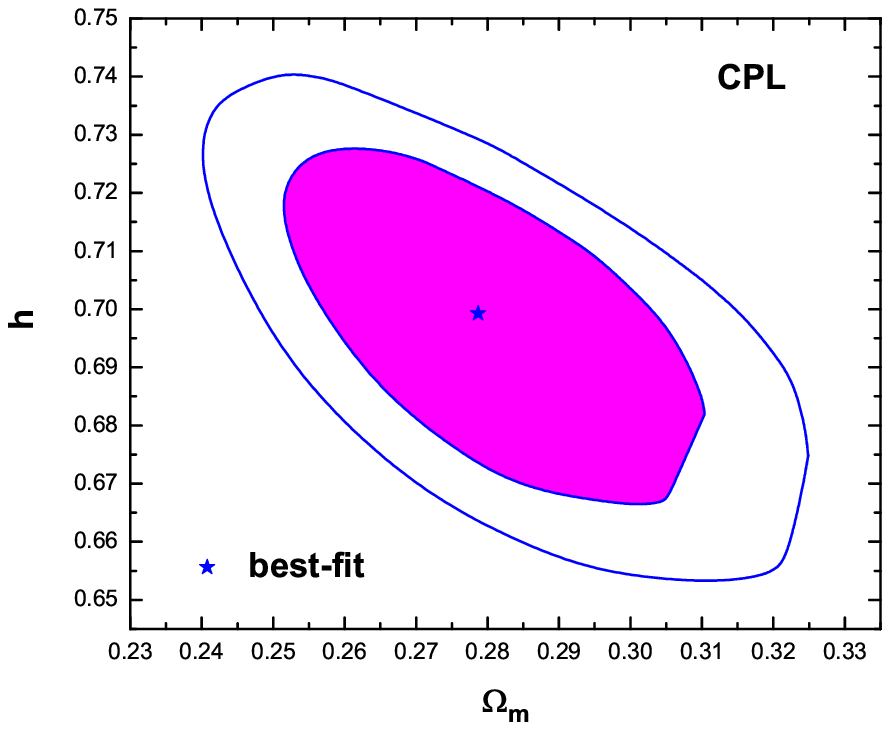} \\
\mbox{(a)} & \mbox{(b)}
\end{array}$
$\begin{array}{c@{\hspace{0.2in}}c} \multicolumn{1}{l}{\mbox{}} &
\multicolumn{1}{l}{\mbox{}} \\
\includegraphics[scale=0.68]{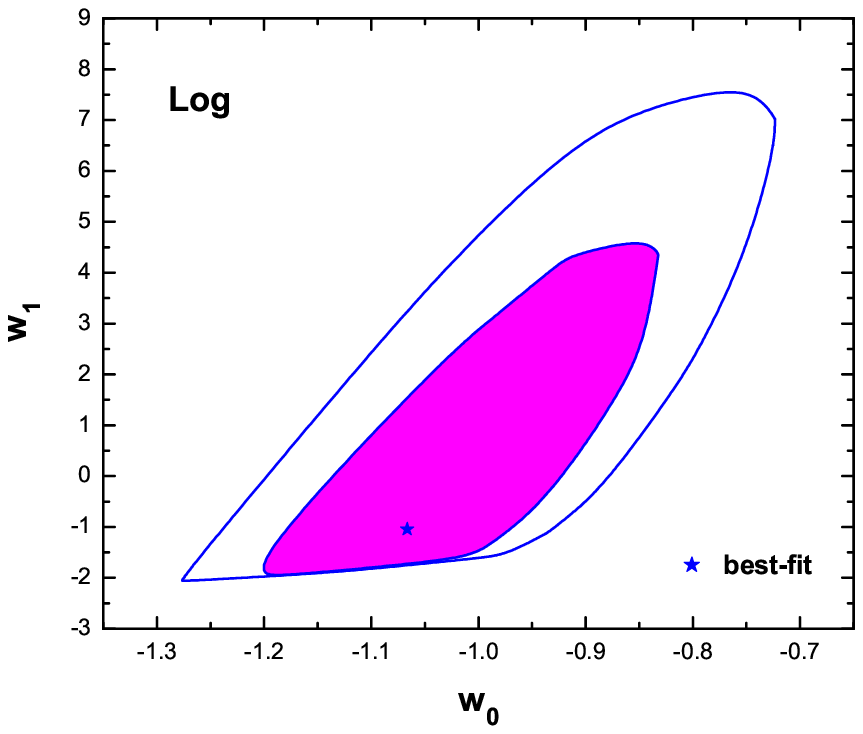} &\includegraphics[scale=0.68]{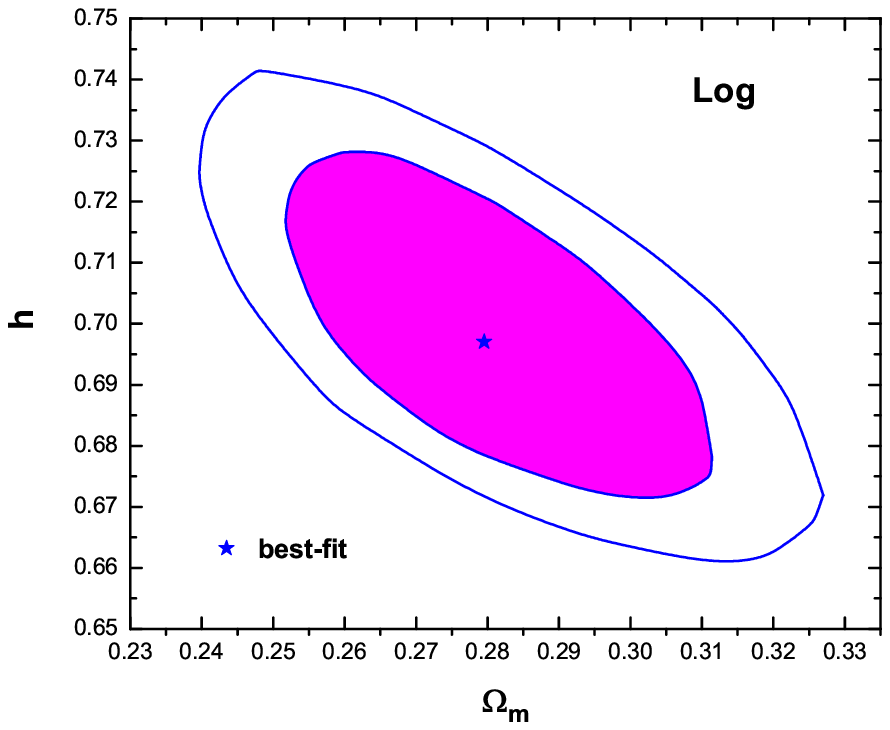} \\
\mbox{(c)} & \mbox{(d)}
\end{array}$
$\begin{array}{c@{\hspace{0.2in}}c} \multicolumn{1}{l}{\mbox{}} &
\multicolumn{1}{l}{\mbox{}} \\
\includegraphics[scale=0.68]{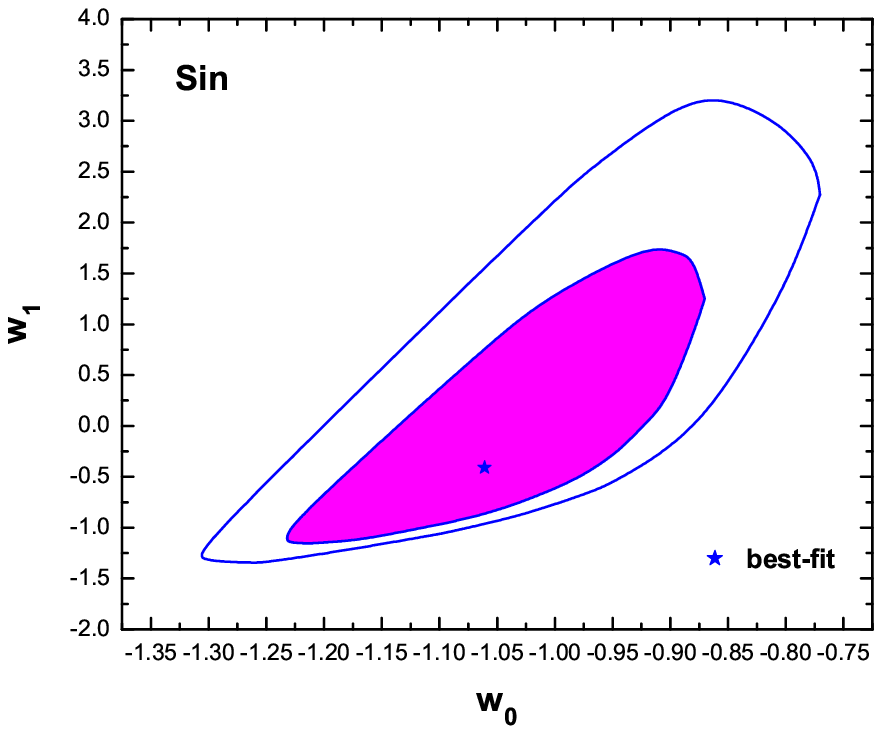} &\includegraphics[scale=0.68]{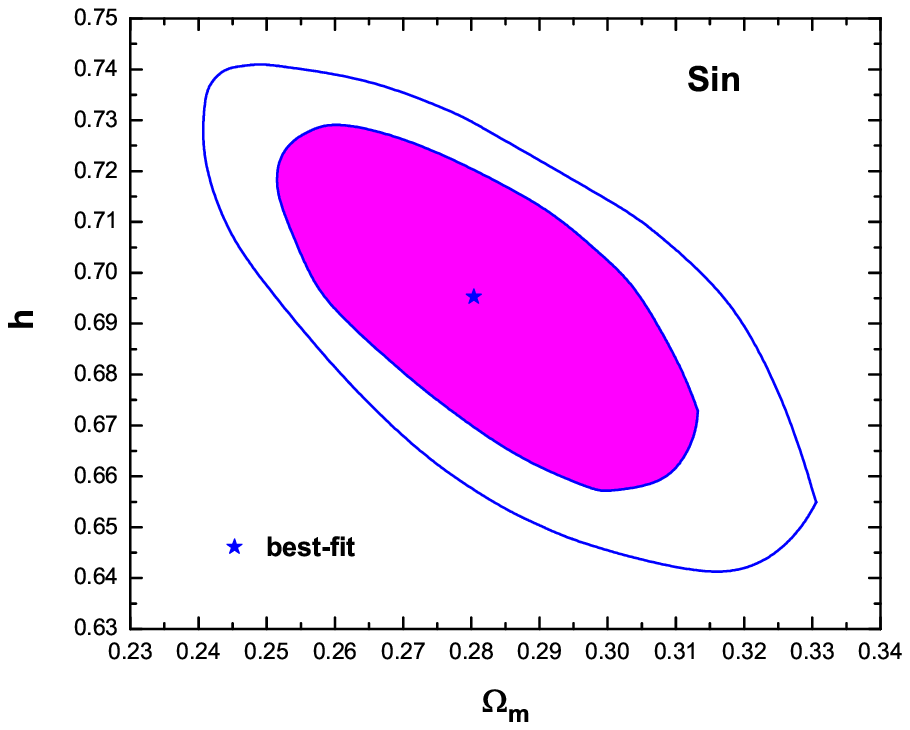} \\
\mbox{(e)} & \mbox{(f)}
\end{array}$
\end{center}
\caption[]{\small \label{fig1}The probability contours at $1\sigma$
and $2\sigma$ confidence levels in the $w_0-w_1$ and $\Omega_{m}-h$
planes for the CPL, logarithm (Log) and oscillating (Sin) models.}
\end{figure*}

\begin{figure*}[htbp]
\centering
\begin{center}
$\begin{array}{c@{\hspace{0.2in}}c} \multicolumn{1}{l}{\mbox{}} &
\multicolumn{1}{l}{\mbox{}} \\
\includegraphics[scale=0.68]{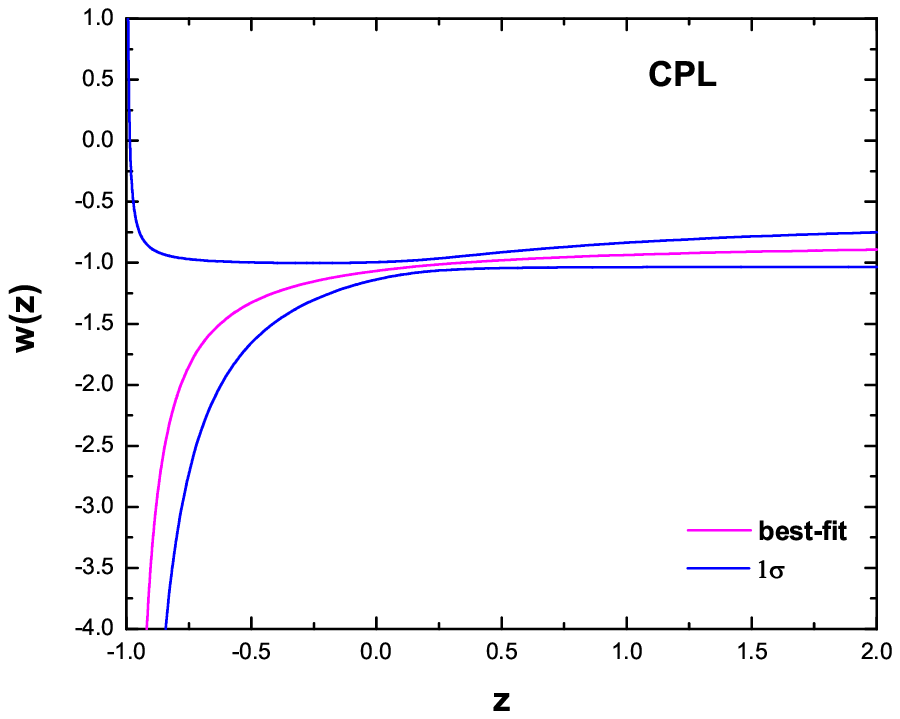} &\includegraphics[scale=0.68]{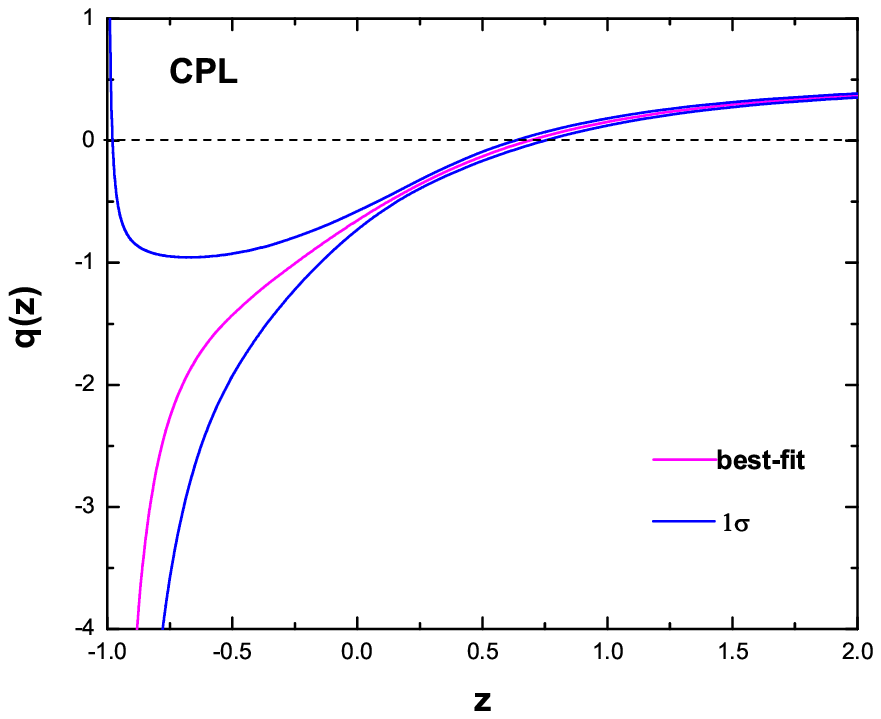} \\
\mbox{(a)} & \mbox{(b)}
\end{array}$
$\begin{array}{c@{\hspace{0.2in}}c} \multicolumn{1}{l}{\mbox{}} &
\multicolumn{1}{l}{\mbox{}} \\
\includegraphics[scale=0.68]{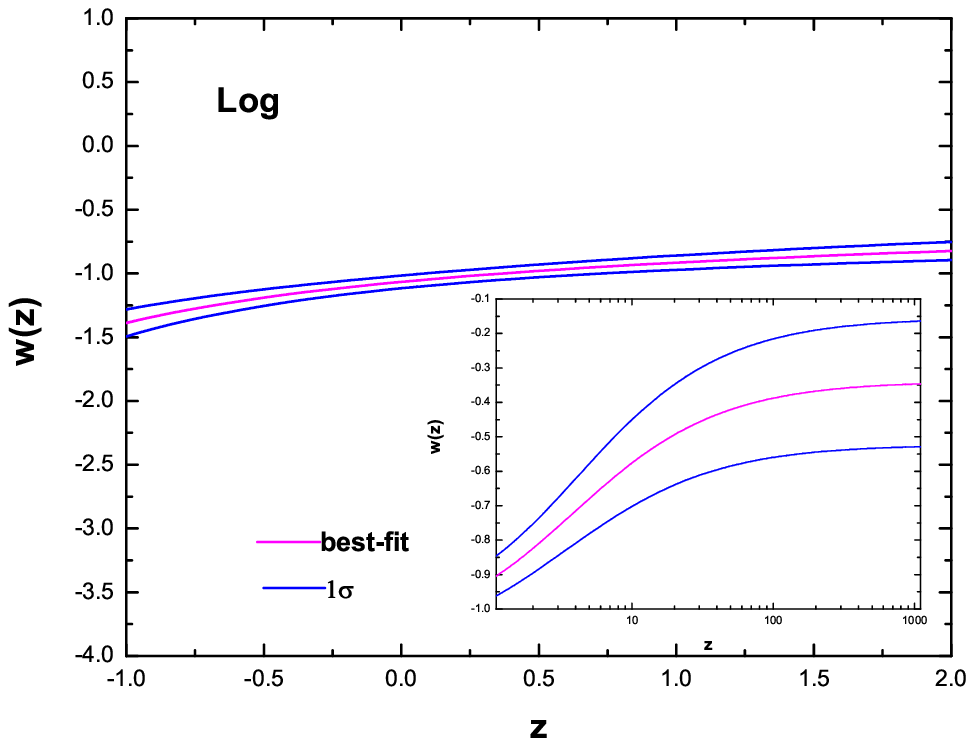} &\includegraphics[scale=0.68]{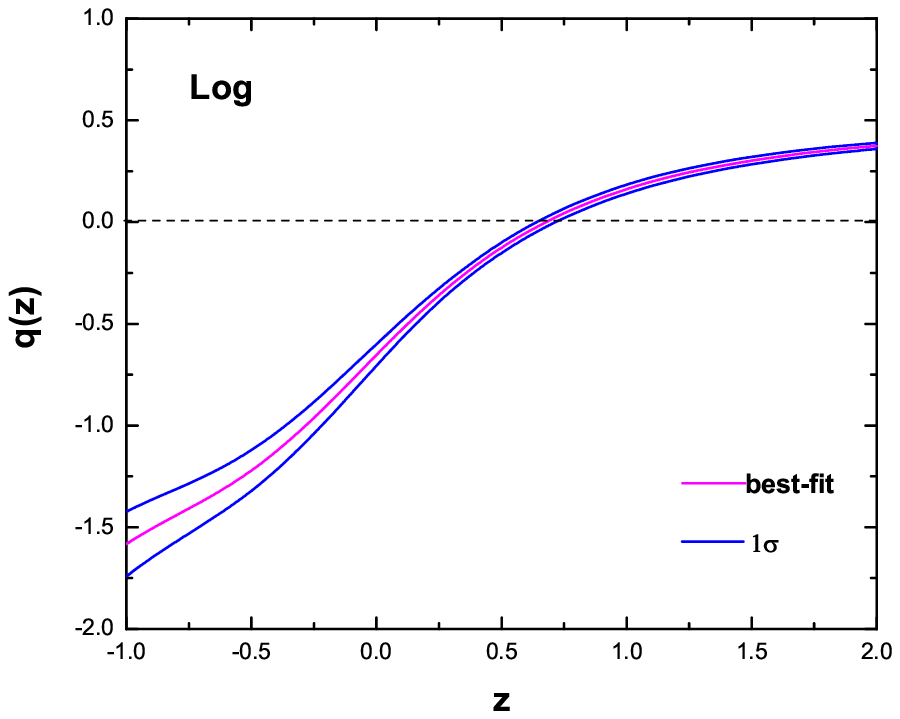} \\
\mbox{(c)} & \mbox{(d)}
\end{array}$
$\begin{array}{c@{\hspace{0.2in}}c} \multicolumn{1}{l}{\mbox{}} &
\multicolumn{1}{l}{\mbox{}} \\
\includegraphics[scale=0.68]{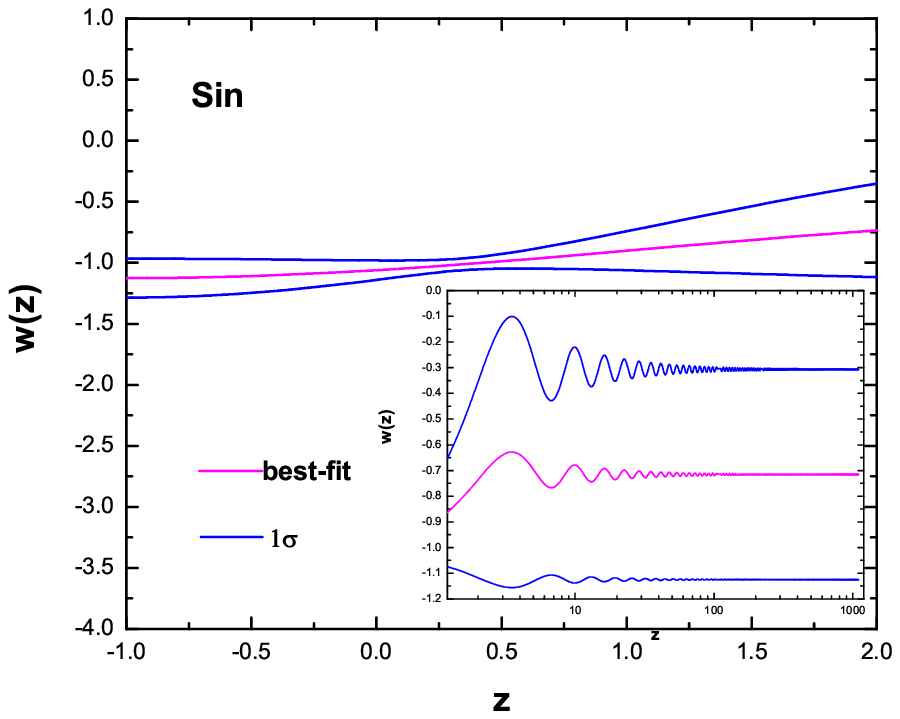} &\includegraphics[scale=0.68]{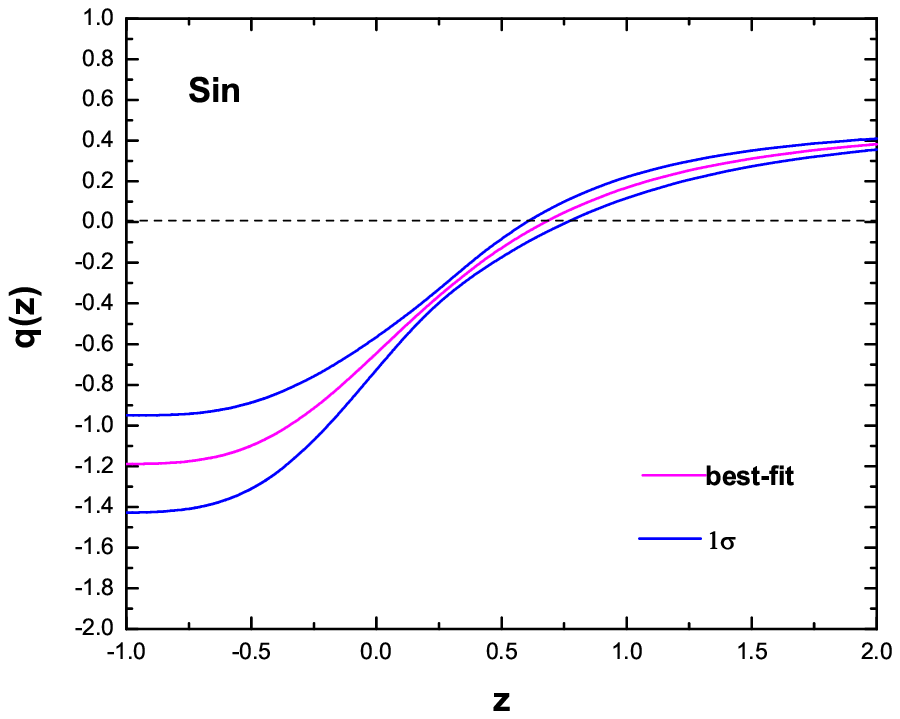} \\
\mbox{(e)} & \mbox{(f)}
\end{array}$
\end{center}
\caption[]{\small \label{fig2} The reconstructed evolutionary
histories (from past to future) for $w(z)$ and $q(z)$ in the three
models.}
\end{figure*}

Figure~\ref{fig1} shows the likelihood contours for the three models
in the $w_0-w_1$ and $\Omega_{m}-h$ planes. For the CPL
parametrization, we have $\Omega_{m}=0.279$, $w_0=-1.066$,
$w_1=0.261$ and $h=0.699$, with $\chi^2_{min}$=544.186. We plot the
likelihood contours for the CPL model in the panels (a) and (b) of
Fig.~\ref{fig1}. For the logarithm parametrization, the fitting
results are $\Omega_{m}=0.280$, $w_0=-1.067$, $w_1=-1.049$ and
$h=0.697$, with $\chi^2_{min}=544.081$, smaller than that of the CPL
model. The likelihood contours for this case are shown in the panels
(c) and (d) of Fig.~\ref{fig1}. For the oscillating parametrization,
we obtain the fitting results: $\Omega_{m}=0.280$, $w_0=-1.061$,
$w_1=-0.410$ and $h=0.695$, with $\chi^2_{min}=543.986$, which is
the smallest of the three. The likelihood contours for this case are
shown in the panels (e) and (f) of Fig.~\ref{fig1}. According to the
$\chi^2_{min}$, the new parametrizations are indicated to be more
favored by the observational data.

Next, we reconstruct the expansion history of the universe in light
of the above fitting results for the three models, and then make a
comparison for them. Since we focus on the properties of dark
energy, we only reconstruct the evolutionary behaviors of the EOS of
dark energy, $w(z)$, and the deceleration parameter of the universe,
$q(z)$. The reconstructing results are shown in Fig.~\ref{fig2}. As
has been pointed out in the above, the CPL model has a problem:
$w(z)$ diverges when $z$ approaches $-1$. So, the CPL
parametrization can only properly describe the past evolution
history but cannot genuinely depict the future evolution; it is
incomplete in describing the evolutionary history of dark energy.
Consequently, the CPL parametrization is not capable of covering
other dark-energy theoretical models. Such a problem can be
explicitly seen in the reconstructed evolution plots, panels (a) and
(b) of Fig.~\ref{fig2}. We can see from these two plots, $w(z)$ and
$q(z)$, that although the CPL model can do a good job in describing
the past evolution of dark energy, it totally loses the prediction
capability for the future evolution of dark energy. The novel
parametrizations successfully overcome the shortcoming of the CPL
model. The reconstructed evolutionary plots, panels (c) and (d) for
the logarithm form (\ref{eq6}) and panels (e) and (f) for the
oscillating form (\ref{eq9}), indicate that the both new models can
nicely describe the whole evolution history of dark energy.

Comparing the panels (a), (c) and (e) of Fig.~\ref{fig2}, we find
that all the three models favor a quintom behavior~\cite{quintom}
that the EOS crosses $-1$ around the recent epoch. This $w=-1$
crossing feature is only mildly favored by the CPL and the
oscillating models, but is explicitly favored by the logarithm
model. For the CPL model, since it forfeits the prediction
capability for the future evolution, we cannot say anything about
the ultimate fate of the universe. For the oscillating model, the
fate of the universe is not definitely determined: the big rip may
or may not occur. For the logarithm model, the universe will
definitely move towards its tragic destiny: the cosmic doomsday will
happen at about 3$\sigma$ level. If we only focus on the past
evolution, we find that the CPL model performs better than the
oscillating model. The best one among the three is without doubt the
logarithm model, not only for the description of the past evolution
but also of the future evolution. From the panels (c) and (e) we see
clearly that the oscillating model behaves similarly to the
logarithm within the range from $z=-1$ to $z=2$, while the
oscillating feature emerges in the oscillating model from a long
term perspective. Comparing the panels of (b), (d) and (f) of
Fig.~\ref{fig2}, we also find that the logarithm model performs the
best. From the reconstructed $q(z)$ plots we see that the
accelerated expansion of the universe starts at a redshift around
$0.5-0.7$. According to the new models, for the future evolution,
the change rate of the cosmic acceleration, $|dq(z)/dz|$, will first
increase and then decrease, with the pivot around $z\approx -0.45$.
We find that for the far future, the change rate $|dq(z)/dz|$ for
the logarithm model is still rather large but for the oscillating
model approaches zero.

In summary, we have proposed two novel parametrizations for the EOS
of dark energy, successfully avoiding the future divergency problem
of the CPL parametrization, and used them to probe the dynamics of
dark energy not only in the past evolution but also in the future
evolution. We pointed out that the CPL parametrization can only
properly describe the past evolution history of dark energy but
cannot genuinely depict the future evolution of dark energy owing to
the divergency of $w(z)$ as $z$ approaches $-1$. Such a divergency
feature forces the CPL parametrization to lose its prediction
capability for the fate of the universe and to fail in providing a
complete evolution history for the dark energy. Consequently, the
CPL model cannot genuinely cover scalar-field models as well as
other dark energy theoretical models. Our new proposals, the
logarithm form $w(z)=w_0+w_1({\ln (2+z)\over 1+z}-\ln2)$ and the
oscillating form $w(z)=w_0+w_1({\sin(1+z)\over 1+z}-\sin(1))$,
exhibit well-behaved features for the EOS of dark energy in all the
evolution stages of the universe. We constrained the new models by
using the current observational data including the 557 Union2 SN
data, BAO data from SDSS DR7 and CMB data from 7-yr WMAP. The
fitting results show that the oscillating parametrization gains a
minimal $\chi^2_{min}$ among the three models, seemingly more
favored by the data. However, by reconstructing the whole
evolutionary histories of $w(z)$ and $q(z)$ from past to future via
the fitting results, we found that the logarithm parametrization is
more tightly constrained by the data. The reconstruction results
show that the $w=-1$ crossing feature is favored. Furthermore, the
new models predict that in the future the cosmic acceleration will
first speed up and then slow down. In the logarithm model, the
cosmic doomsday seems inevitable, which will happen at about
$3\sigma$ level. We believe that the novel parametrizations proposed
in the present Letter deserve further investigations.


\begin{acknowledgments}
We thank Yun-He Li and Jing-Fei Zhang for useful discussions. This
work was supported in part by the Natural Science Foundation of
China under Grant Nos.~10705041 and 10975032, as well as the
National Ministry of Education of China under the innovation program
for undergraduate students.
\end{acknowledgments}


\end{document}